# INSPECTION OF METHODS OF EMPIRICAL MODE DECOMPOSITION

Roberto Hernández Santander[1] and Esperanza Camargo Casallas, PhD[2]

[1]Faculty of Engineering, Universidad Distrital Francisco José de Caldas, Bogotá, Colombia
rshernandezs@correo.udistrital.edu.co

[2] Faculty of Technology, Universidad Distrital Francisco José de Caldas, Bogotá,
ecamargoc@udistrital.edu.co

*ABSTRACT*

*Empirical Mode Decomposition is an adaptive and local tool that extracts underlying analytical components of a non-linear and non-stationary process, in turn, is the basis of Hilbert Huang transform, however, there are problems such as interfering modes or ensuring the orthogonality of decomposition. Three variants of the algorithm are evaluated, with different experimental parameters and on a set of 10 time series obtained from surface electromyography. Experimental results show that obtaining low error in reconstruction with the analytical signals obtained from a process is not a valid characteristic to ensure that the purpose of decomposition has been fulfilled (physical significance and no interference between modes), in addition, freedom must be generated in the iterative processes of decomposition so that it has consistency and does not generate biased information. This project was developed within the framework of the research group DIGITI of the Universidad Distrital Francisco José de Caldas.*

## 1. INTRODUCTION

The analysis of data always serves to generate a simple model that through different characteristics can represent a phenomenon, and the main problems are evoked due to the insufficiency in the quantity of data and that these represent non-linear and/or non-stationary dynamics. Several methods exist, and to a different extent restricted to periodic and stationary characters, and none of them can guarantee the representation of a non-linear and non-stationary signal with a base for complete, orthogonal, local and adaptive conditions, something that the Hilbert-Huang transform can do, which is an emerging technique that offers the possibility of performing a multi-scale, high-resolution, time-frequency-power analysis [1]. The requirement of the transformed is the acquisition of analytical signals, those whose frequency spectrum is null for negative frequencies and with real part equal to the original signal, therefore, of such topology that they allow the definition of instantaneous frequency in all the temporal distribution. For this reason, EMD becomes important, since intrinsic mode functions are obtained, IMF, components that together reconstruct the study phenomenon, representing underlying dynamics that are individually represented as appropriate analytical signals for the application of the Hilbert transform and its subsequent study of frequency and instantaneous power through the Hilbert-Huang transform [2]. It is doubtful to find a way to apply EMD over a time series, such that it can reconstruct the signal properly, and also really represents the underlying physical significance of the phenomenon studied.

Since the publication of Huang [1], several studies have focused on the application and improvement of EMD, generating new stop criteria, as Rilling [3], in which a criterion is defined that is not based on two thresholds, local and global, to consider variations of the mean at a different scale, contrary to that which is based on a measure of deviation of two consecutive iterations as stated [1]. Then, Huang [4] establishes as a stop criterion, to brake when the IMF conditions posed by [1] are met a certain number of times. Other works propose variations in which the signal is directly affected. Deering and Kaiser [6], generate a method of masking the signal with pairs of complementary noise that once they have established a consistent background in the spacetime frequency for decomposition, are easily eliminated as they are anti-correlated. On the other hand, Wu and Huang [7], propose the EEMD method, where also the properties of the noise are used, this time used a certain amount of realizations such that in the end the false information is eliminated through an average between the IMFs of each realization. Finally, Flandrin [8], with the CEEMDAN method suggests to change the dynamics of the decomposition algorithm, using noise as [7] but getting better results in terms of mixing modes, and rebuilding errors.

Beyond the number of possible variants of EMD, the real problem with dealing with this work is that there really is no general standard set on how the decomposition algorithm should be implemented. Many jobs have implemented EMD, all work with different types of time series. To mention a few, [11] seeks to analyse earthquake signals thanks to the energy distribution of the temporal frequency given the EMD method proposed by Huang [1] with the HHT. For its part [12] identifies six emotional states through ECG signals using the same criteria, and in the same way [13] to analyse heart rate variability, [14] to do a time frequency analysis of ECG signals, and [15] and [16] to identify muscle activation characteristics in myoelectric signals and voice recognition respectively. All of the above reach favourable results in their investigations, as new components with analysable information are effectively obtained through decomposition; however, the importance of mitigating the mix of modes to ensure exclusive information between components in a way that ensures physical significance is left aside. Works like [17] and [18], focus more on this fact, implementing the first one a variant of the method exposed by [6] for EEG signals, where noise is used to avoid the mixing of modes, and the second one using the EEMD method proposed by [7] to analyse EEG signals.

Given the above, it is extremely important to study how to ensure that each IMF presents relevant and unique information, being consistent with the fact that this surely depends on the topology of the signal to be treated. The present work reviews the studies carried out on decomposition in an empirical way and the main variants that have been carried out by research, with these, we experiment using 10 time series of different wrist movements, acquired with superficial electromyography in the forearm, with the aim of establishing a framework on the general behaviour of the different decomposition models, and determine under which variation the best results are obtained.

## 2. EMD VARIATIONS

### 2.1. Standard EMD

EMD [1] considers the oscillations of signals at a local level, in order to find functions intrinsically. An intrinsic mode function (IMF) is a function that satisfies two conditions: (1) in the whole data set, the number of extrema and the number of zero crossings must either equal or differ at most by one; and (2) at any point, the mean value of the envelope defined by the local maxima and the envelope defined by the local minima is zero. These are found by extracting from each oscillation the local detail d(t) and the local trend m(t). Given an x(t) signal, the EMD algorithm is summarized as follows [1][3]:

1. Identify all extrema of x(t) and interpolate between minima ending up with some envelope emin(t). To do the same thing with the maxima to find emax(t)
2. Compute the mean m(t)= (emin(t)+emax(t))/2
3. Extract the detail d(t)= x(t)-m(t)

4. Iterate steps 1 to 4 upon the detail signal d(t), until this latter can be considered as zero-mean according to some stopping criterion, then d(t) satisfy the definition of IMF.
5. If does not satisfy the definition, repeat step 1 to 5 on d(t) as many times as needed till i satisfies the definition
6. If satisfy the definition, assign d(t) as an IMF component, c(t)
7. Repeat the operation step 1 to 7 on the residue, r(t)=x(t)-c(t). The operation ends when the residue contains no more than one extremum

The variants for stop criteria to study about this method are:
- EMD 1: Disable the default stopping criterion and do a number N of sifting iterations with |#zeros-#extrema|<=1 to stop [4].
- EMD 2: Disable the default stopping criterion an do exactly a number N of sifting iterations for each mode.
- EMD 3: Huang [1] determines the SD standard deviation criterion, calculated from two consecutive iteration results, whose typical value can be set between 0.2 and 0.3.
- EMD 4: Rilling [3] introduce a criterion based on 2 thresholds $\theta 1$ and $\theta 2$, aimed at guaranteeing *globally* small fluctuations in the mean while taking into account *locally* large excursions. This amounts to introduce the *mode amplitude* $a(t) := (emax(t) - emin(t))/2$ and the *evaluation function* $\sigma(t) := |m(t)/a(t)|$ so that sifting is iterated until $\sigma(t) < \theta 1$ for some prescribed fraction $(1 - \alpha)$ of the total duration, while $\sigma(t) < \theta 2$ for the remaining fraction. One can typically set $\alpha \approx 0.05$, $\theta 1 \approx 0.05$ and $\theta 2 \approx 10\, \theta 1$.

## 2.2. EEMD

Ensemble Empirical Mode Decomposition defines IMF components as the average of IMFs obtained through EMD performed several times, each with finite variance white noise added to the x(n) signal. The algorithm is [7],[8]:

1. Generate $x^i(n) = x(n) + w^i(n)$, where $w^i(n)$ are different realizations of White Gaussian noise
2. Each $x^i(n)$ is fully decomposed by EMD getting their modes $IMF_k^i(n)$, where $k = 1, ..., K$ indicates the modes,
3. Assign $\overline{IMF\_k}$ as the $k - th$ mode of $x(n)$, obtained as the average of the corresponding $IMF_k^i$: $\overline{IMF_k}(n) = \frac{1}{I}\sum_{i=1}^{I} IMF_k^i(n)$

The uniformly distributed noise, assembled with the time series, as exposed [9], generates for EMD a diadic filter behavior, generating more effective and isolated IMF components in terms of their physical processes, this because the noise provides a uniformly distributed scale in the spacetime frequency that is associated with the intrinsic oscillations of the signal at different scales [10]. For this reason, in principle increasing the standard deviation of the noise should be beneficial, however, this is not trivial. In the first instance, since this is not correlated, it can be eliminated through a series of averages between different realizations of the same signal with different noise signals, however the noise could be associated intrinsically, or a very large number of realizations could be required to eliminate the noise from the components [8],[9],[10].

In order to analyse what consequences can be developed from the noise to mitigate the mode crossover, and the method's ability to eliminate residual noise from the components, the variant parameters for this case are the noise standard deviation *Nstd*, and the number of realizations *NR*, on a pre-established EMD model.

## 2.3. CEEMDAN

Complete Ensemble Empirical Mode decomposition with adaptive noise [8] is a method in which, contrary to EEMD, each $x^i(n)$ does not decompose independently. In the method presented there, the decomposition of modes is calculated in such a way that a first residue is obtained from the mean of the first IMF, and the process continues successively. The decomposition is done as follows:

1. Decompose with EMD, N realizations: $x(n) + \varepsilon_0 w^i(n)$ to get the first mode $\widetilde{IMF_1}(n) = \overline{IMF_1}(n)$.
2. Calculate the first residue: $r_1(n) = x(n) - \widetilde{IMF_1}(n)$
3. Decompose $r_1(n) + \varepsilon_1 E_1 w^i(n)$, $i = 1, ..., N$, until you find the first EMD mode and thus define the second CEEMDAN mode: $\widetilde{IMF_2}(n) = \frac{1}{N}\sum_{i=1}^{N} E_1(r_1(n) + \varepsilon_1 E_1 w^i(n))$
4. For $k = 2,, .....K$ calculate el k-th residue: $r_k(n) = r_{(k-1)}(n) - \widetilde{IMF_k}(n)$
5. Decompose realizations $r_k(n) + \varepsilon_k E_k w^i(n)$, $i = 1, ..., N$ until their first EMD mode and define $(k + 1)$-th mode as $\widetilde{IMF_{(k+1)}}(n) = \frac{1}{N}\sum_{i=1}^{N} E_1(r_k(n) + \varepsilon_k E_k w^i(n))$
6. Go to step 4 for next k.

Steps 4 to 6 are performed until the obtained residue is no longer feasible to be decomposed. As in the EEMD case, the variant parameters are *Nstd* and *NR*, on a preset EMD model.

## 3. ANALYSIS DATA

The set of test signals used is shown in Figure 1. Each signal is sampled at a different frequency, between 600 and 700 Hz, and corresponds to biological measurements, taken in the superficial flexor muscle of the fingers and the ulnar flexor muscle of the carpus for five different movements. To evaluate the results an evoked analysis is made to search for interference between modes, and to study the reconstruction of the signal through the generated IMFs, the reconstruction of the 10-signal test set is evaluated by finding the average quadratic mean error of all-time series.

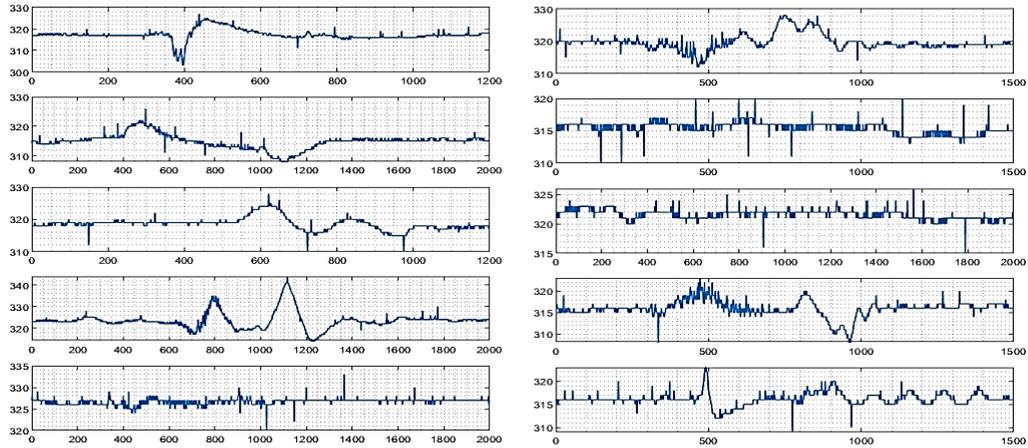

Figure 1. Test time series

Each time series in Figure 1 represents a biological phenomenon. These are considered to develop the study for possessing in general characteristics of low and high frequency, and in addition, each of these is an assembly of different properties, all dependent on the musculature and processes between them, as well as the physiology and anatomy of each individual, facts that in principle can be extracted and analysed individually with EMD. These time series are experimented on using the three variants of the decomposition algorithm expressed above, in order to perform a differential analysis of the variants of the algorithm tested since its standard edition [1], the variants posed by [2] and [5] and the background modifications that occur due to the contamination of the study signals and the modification of the way in which the MFIs are extracted, as proposed by the EEMD method and CEEMDAN.

It should be noted that this study is limited to the topology of the study signals used, a fact that must be taken into account each time an EMD method is performed, therefore it is assumed that all results are suitable for time series concerning superficial forearm electromyography, aperiodic, non-linear, non-stationary and for those with similar temporal and frequency characteristics.

## 4. EXPERIMENTATION AND RESULTS

### 4.1. Standard EMD experimentation

Six experiments were performed with the EMD algorithm as specified by [1] considering the variants regarding the stop criterion implemented by [2] and [5]. The experimental parameters and results are summarized in Table 1.

Table 1. Results of EMD experimentation.

| Exp. | Variant | Parameters | IMF | Time (s) | Iterations | ECM |
|---|---|---|---|---|---|---|
| 1 | EMD 1 | N=10 | 10 | 0.63 | 759 | 0.08 |
| 2 | EMD 2 | N=10 | 8 | 0.04 | 73 | 0.96 |
| 3 | EMD 3 | SD=0.2 | 10 | 68.1 | 542 | 0.25 |
| 4 | EMD 4 | $\alpha \approx 0.05, \theta1 \approx 0.05, \theta2 \approx 0.5$ | 8 | 0.05 | 63 | 0.49 |
| 5 | EMD 4 | $\alpha \approx 0.25, \theta1 \approx 0.25, \theta2 \approx 2.5$ | 8 | 0.05 | 73 | 0.95 |
| 6 | EMD 4 | $\alpha \approx 0.01, \theta1 \approx 0.01, \theta2 \approx 0.1$ | 8 | 0.04 | 71 | 0.83 |

The parameters of the first two experiments allude to the characteristic of diadic filters [5] that decomposition can acquire by establishing 10 as the maximum number of iterations per screening process to maintain a balance between separation between components and information leakage. The third experiment performs its function as Huang puts it [1], in which it is iterated until there is no local zone with considerable mean. Finally, the last three experiments, based on a less narrow criterion with two measures of mean, local and global [3], are analysed in such a way as to intuit the behavior that they take to different measures of strictness.

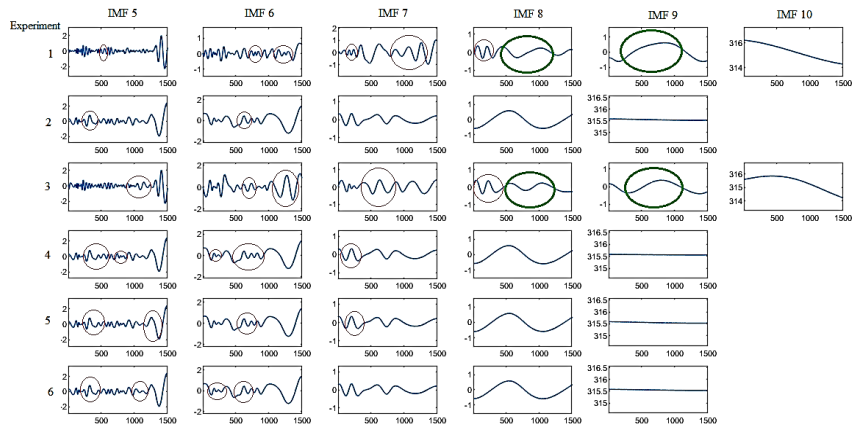

Figure 2. last IMFs of the six EMD experiments

The results in Figure 2 are obtained by the fourth time series in Figure 1. As shown, there is no significant difference in the results of experiments 2, 4, 5, and 6. Due to the emergence of one more IMFs for experiments 1 and 3, less mix of modes would be expected, however, this phenomenon is still observable. Figure 2 indicates with thin circles the mix of modes, and with pronounced circles, the difference between decompositions of MFIs 8 and 9. A different level of detail is observed, proving superiority with experiment 3, but, although methods 1 and 3 have the least error, method four is the one that best achieves a balance between avoiding the mix of modes and having low error levels.

### 4.2. EEMD Experimentation

In this case, variations of *Nstd* and *NR* are made. The idea is to observe what happens when a noise level is maintained and the number of realizations varies, and what happens when experimentation is given to the contrary, varying the noise and maintaining the number of realizations, evaluating also low and high values in general. Six experiments are performed with the parameters and results of table 2. The basic EMD algorithm is adjusted to operate under the criterion of experiment 4 of EMD, with a maximum of iterations *MaxIter=10*.

Table 2. EEMD Experimentation.

| Exp. | NR    | Nstd | IMF | Time (s) | Iterations |
|------|-------|------|-----|----------|------------|
| 1    | 500   | 0.02 | 11  | 25.42    | 36801      |
| 2    | 2000  | 0.02 | 11  | 100.36   | 147441     |
| 3    | 10000 | 0.02 | 11  | 492.36   | 736774     |
| 4    | 1000  | 0.05 | 11  | 61.15    | 75561      |
| 5    | 1000  | 0.1  | 11  | 51.26    | 75657      |
| 6    | 1000  | 0.5  | 11  | 50.3     | 75406      |

The results obtained are shown in figure 3. The increase in NR of experiments 1-3 reduces the ECM, an effect that is contrasted with experiments 4-6, with which it is noted that the increase in noise generates an accelerated growth of the error. With this experimentation it is found that, to a certain extent, *Nstd* levels can be counteracted with the increase of NR, however, it does not favour considerably the decrease of the error, therefore, to arrive at minimum errors close to those obtained under EMD, and to eliminate the totality of the noise, NR should tend to infinity, in addition the most dramatic effect of corruption in the results is due to the increase of *Nstd*, and to a lesser extent to low values of NR.

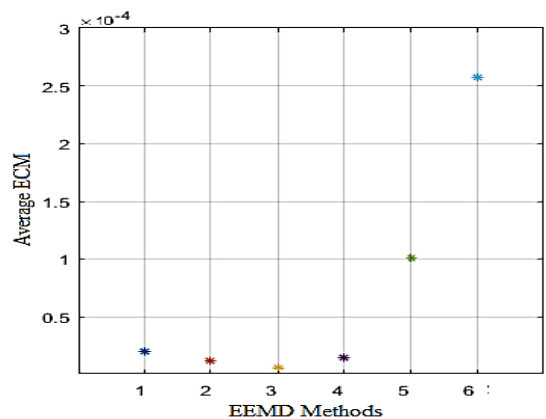

Figure 3. Experimental errors EEMD

### 4.3. CEEMDAN Experimentation

A total of 32 experiments are performed on the method. The first six under the characteristics of table 2 of the EEMD experiment. As can be seen, for the first six experiments very low errors are obtained compared to the EEMD method, but the process is indifferent to the variations of *Nstd* and *NR*. The experimentation is extended in order to study the implication of each variant parameter. Table 3 shows the computational expenditure of the CEEMDAN method, which despite being used as the EEMD method, generates for all cases greater expenditure of resources.

Table 3. CEEMDAN Experimentation.

| NR | Nstd | IMF | Time(s) | Iterations |
|---|---|---|---|---|
| 500 | 0.02 | 10 | 75,89 | 42181 |
| 2000 | 0.02 | 10 | 252,53 | 170122 |
| 10000 | 0.02 | 10 | 1217,21 | 851710 |
| 1000 | 0.05 | 11 | 116,28 | 85599 |
| 1000 | 0.1 | 10 | 111,53 | 80096 |
| 1000 | 0.5 | 11 | 112,95 | 79122 |

Subsequently, 8 more experiments were performed by varying *Nstd* maintaining *NR=500* and MaxIter=10, with which contradictory results are obtained. Figure 5 shows low errors for the highest noise levels, caused by the fact that the high noise levels are filtered by the average of components to eliminate the noise between the realizations of the CEEMDAN method, contrary to what happens with low noise levels, which, given the topology of the signals, enter a state in which they unify the components, reaching correlation states that limit the finding of unbiased IMFs. Even so, the mix of modes is mitigated by increased noise as shown in Figure 4.

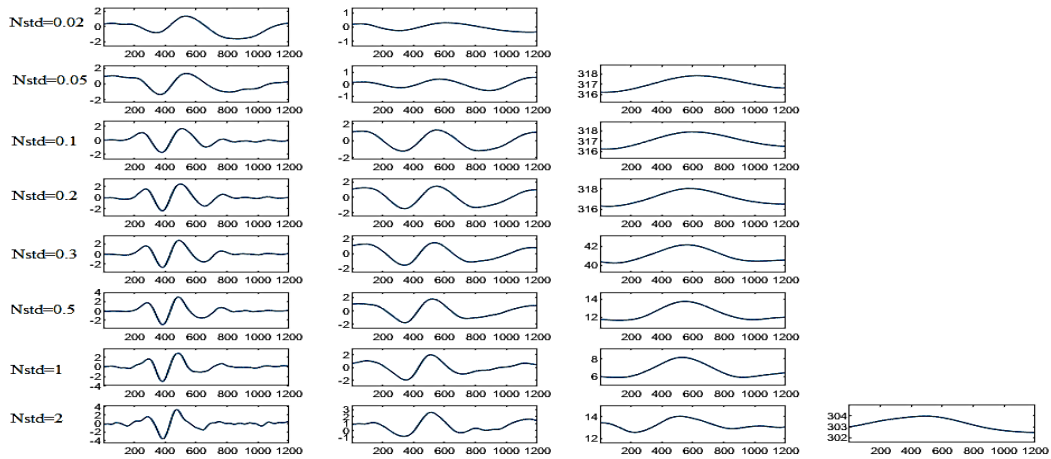

Figure 4. Latest IMF under CEEMDAN method with noise variation. *NR=500*

With this experiment it is noticeable that low noise values cause low level of detail and mix of modes. When *Nstd* takes a high value, the number of extremes in the functions increases, which generates more details in the modes, implying that the frequency in the forms tends to increase and more modes appear to compensate for lower frequencies that start to get lost. Considering this and the above it is determined that the most correct option should be a central *Nstd* value equal to *0.2*, even though this option takes longer than the other methods (see table 4).

Table 4 shows that under these experimental parameters there is an increase in the number of IMFs and in addition, although time remains around 74s, the number of iterations tends to decrease with the increase in noise, this means that the last information lags corresponding to the last IMFs tend to have a more regular harmonic behavior with the increase in noise, which consequently leads to information with more detail easily extractable and at the same time, more biased and with a greater possibility of falling into a mix between modes.

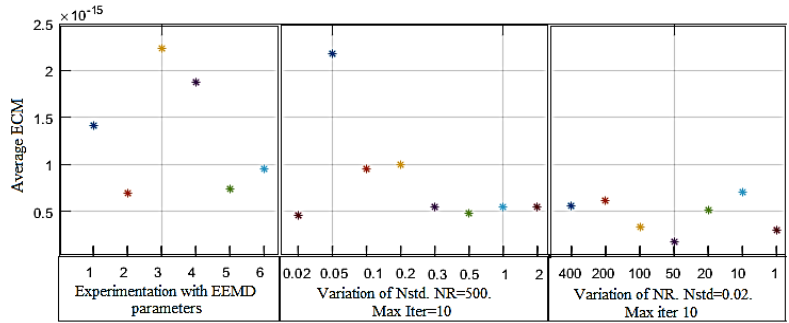

Figure 5. Errors in experimentation with the CEEMDAN method

Table 4. Results of CEEMDAN experimentation varying noise. NR=500

| Nstd | IMF | Time (s) | Iterations |
|---|---|---|---|
| 0.02 | 10 | 66,72 | 42299 |
| 0.05 | 11 | 77,92 | 43276 |
| 0.1 | 11 | 72,64 | 42437 |
| 0.2 | 11 | 81,51 | 40244 |
| 0.3 | 11 | 69,33 | 38937 |
| 0.5 | 11 | 59,76 | 35909 |
| 1 | 11 | 72,94 | 30998 |
| 2 | 12 | 70,89 | 27767 |

To check the state of corruption into which signals can fall when contaminated with low levels of noise that mix indistinctly with real information, seven more experiments are performed establishing Nstd=0.02, maintaining MaxIter=10 and varying *NR*. The results show (see figure 5) that the changes are not apparently related to an optimum in the results, therefore, under these characteristics, the effects that are expected to be achieved with the method have been mitigated, since as it is observed, it has almost the same effect to carry out a realization or 400.

Table 5. Results of experimentation CEEMDAN varying *NR*. Nstd=0.02

| NR | IMF | Time (s) | Iterations |
|---|---|---|---|
| 400 | 10 | 53,35 | 33916 |
| 200 | 10 | 30.50 | 16685 |
| 100 | 10 | 14,10 | 8351 |
| 50 | 10 | 6,96 | 4183 |
| 20 | 11 | 3,19 | 1755 |
| 10 | 11 | 1,32 | 884 |
| 1 | 12 | 0,14 | 94 |

Table 5 shows that the *NR* exchange rate corresponds to time and iteration variations, however, IMFs tend to increase, an effect similar to that presented in the experimentation with increased noise level. This means that although the variation of *NR* does not have considerable negative effects on the reconstruction of the signal, in the time-frequency space a mixture is generated between modes and biased information, overlapping values of power and frequency of the real information.

Finally, 10 experiments are carried out with freedom of iterations (*MaxIter=5000*), so as to ensure that the stop criterion established by Rilling [2], under the EMD parameters used in experimentation 4 of EMD. In the first five experiments *Nstd=0.2* is established and *NR* is varied. In the last five experiments *NR=500* is established and *Nstd* is varied. The results in Figure 6 show a differentiation between the variations of the algorithm and a coherent response, with intuitive behaviours that support variety, giving the opportunity to establish consistent parameters for experimentation. It is observed how high noise levels and low number of realizations generate the same effect, increase the error, and in the same way, the error decreases, when NR increases and *Nstd* decreases.

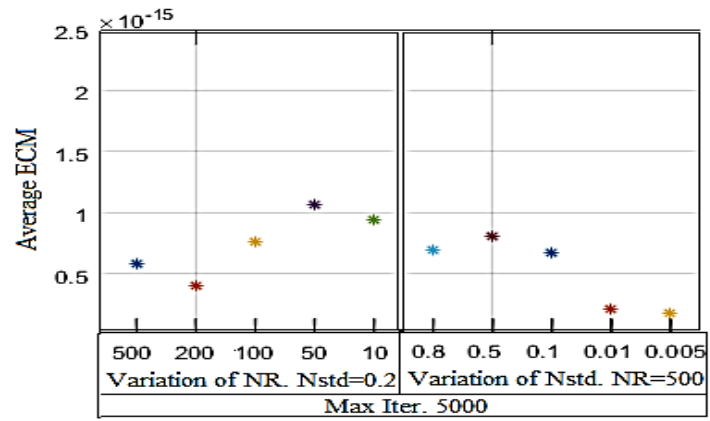

Figure 6. Results of experimentation CEEMDAN extending maximum of iterations

Table 6 shows that this experiment has similar behaviours to those previously shown, increasing the expenditure at low noise levels or, as expected, higher NR.

Table 6. CEEMDAN Experimentation, *MaxIter=5000*

| Nstd | NR | IMF | Time (s) | Iterations |
|---|---|---|---|---|
| 0.2 | 500 | 11 | 89,77 | 67410 |
|  | 200 | 11 | 39,53 | 28564 |
|  | 100 | 11 | 19,67 | 15185 |
|  | 50 | 11 | 11,95 | 8066 |
|  | 1 | 12 | 2,39 | 1923 |
| 0.8 | 500 | 11 | 76,28 | 41488 |
| 0.5 |  | 11 | 78,35 | 48313 |
| 0.1 |  | 11 | 108,11 | 119864 |
| 0.01 |  | 11 | 393,63 | 576008 |
| 0.05 |  | 10 | 327,09 | 486814 |

## 4. CONCLUSIONS

Under any EMD method, the algorithm always seeks to reconstruct the time series effectively, regardless of the variant it will try to achieve a minimum error even if it implies the appearance of more IMFs. This in no way ensures that there is no mixture of modes and therefore the orthogonality of the transformation, determining that the veracity of the information cannot be used as a minimum error in the reconstruction with IMF components, and although EEMD manages to mitigate the mixture of modes, given the high errors, there would never be absolute certainty about the information being handled. In addition, a high or low level of noise either does not mitigate the mode mix, or generates it from spurious oscillations that skew the information, bringing more IMFs. On the other hand, a low *NR* has a similar effect to high *Nstd* cases: on any scale the noise generates misleading oscillations.

It is discovered that the restriction to a low level of iterations by screening process involves the risk of annulling the faculties of the CEEMDAN algorithm, therefore, assuring the property of diadic filters for the case study, corrupts the acquisition of IMFs, leaving the algorithm at the mercy of the randomness of the contamination noise. Under the previously described, the CEEMDAN is established as the preferential method, under *Nstd=0.2, NR=500,* with the fourth EMD variant with *MaxIter=5000*, as a criterion to ensure correct dynamics in decomposition, making it clear that the decomposition demonstrates to depend also on the topology of the study signals.

**Authors**

Roberto Hernández was born in San Juan de Pasto, Colombia. He is currently completing his studies in electronic engineering at the School of Engineering of the Universidad Distrital Francisco José de Caldas, develops research in the areas of signal analysis from the framework of bioengineering. In addition, with projection in the areas of computational intelligence, control, automation and industrial electronics.

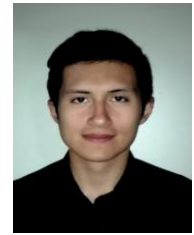

Esperanza Camargo was born in Bogotá, Colombia. Engineer in electronic control and instrumentation of the Universidad Distrital Francisco José de Caldas de Bogotá D.C, Colombia. Specialist in Electronic Instrumentation at Universidad Santo Tomás de Bogotá D.C, Colombia. PhD in engineering from Pontificia Universidad Javeriana, Bogotá D.C, Colombia. Director of the research group DIGITI at the Universidad Distrital Francisco José de Caldas, with research trajectory framed in the areas of Bioengineering and Aerospace Engineering. She is currently the coordinator of the Electronics Technology and Control Engineering and Telecommunications Engineering Curriculum Project at the Universidad Distrital Francisco José de Caldas in Bogotá D.C., Colombia.

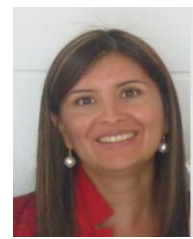